\documentclass[prd,aps,twocolumn,amsfonts,showpacs,superscriptaddress,preprintnumbers,nofootinbib]{revtex4-2}
\usepackage[normalem]{ulem}
\usepackage[utf8]{inputenc}
\usepackage{amsmath,amssymb,amsthm}
\usepackage{dsfont}
\usepackage{physics,mathtools}
\usepackage{tikz,float}
\usepackage{booktabs}
\usepackage[hidelinks]{hyperref} 

\usepackage{siunitx}
\newlength{\lostwd}          
\makeatletter
\renewcommand\section{%
  \@startsection{section}{1}{\z@}%
  {2.2ex plus 0.6ex minus 0.3ex}
  {1.0ex plus 0.2ex}
  {\normalfont\small\bfseries\centering}%
}
\renewcommand\subsection{%
  \@startsection{subsection}{2}{\z@}%
  {1.8ex plus 0.5ex minus 0.3ex}
  {0.8ex plus 0.2ex}
  {\normalfont\small\bfseries\centering}%
}
\renewcommand\subsubsection{%
  \@startsection{subsubsection}{3}{\z@}%
  {1.5ex plus 0.4ex minus 0.3ex}
  {0.7ex plus 0.2ex}
  {\normalfont\small\itshape\centering}%
}
\makeatother

\numberwithin{equation}{section}  
\makeatletter
\def\@fpheader{\vspace{0.1mm}}
\makeatother
\newcommand{\bw}{\begin{widetext}}
\newcommand{\ew}{\end{widetext}}
\newcommand{\bea}{\begin{eqnarray}}
\newcommand{\eea}{\end{eqnarray}}
\newcommand{\be}{\begin{equation}}
\newcommand{\ee}{\end{equation}}
\newcommand{\bca}{\begin{cases}}
\newcommand{\eca}{\end{cases}}

\newcommand{\ben}{\begin{enumerate}}
\newcommand{\een}{\end{enumerate}}

\definecolor{reed}{rgb}{0.6, 0.0, 0.0}
\definecolor{dred}{rgb}{0.6, 0.0, 0.0}

\usetikzlibrary{patterns}
\makeatletter
\let\over\@@over
\makeatother

\makeatletter
\skip\footins=12pt plus 4pt minus 4pt   
\def\footnoterule{%
  \dimen@\skip\footins \divide\dimen@\f@ur   
  \kern-\dimen@\hrule width.5in\kern\dimen@
}
\makeatother

\usepackage[table]{xcolor}
\usepackage{colortbl}
\definecolor{ddred}{RGB}{198,31,31}
\definecolor{ddgreen}{RGB}{0,160,77}
\usepackage{hyperref}
\hypersetup{
    colorlinks=true,
    linkcolor=ddred,
    filecolor=blue,      
    urlcolor=ddred,
    citecolor=ddgreen}
\usepackage{tocloft}

\begin{document}

\title{Where's Kasner? OPE Signature of Kasner Exponents}
\author{Nejc \v{C}eplak}
\affiliation{School of Mathematics and Hamilton Mathematics Institute, Trinity College, Dublin 2, Ireland}
\affiliation{School of Mathematics \& Statistics, University College Dublin, Belfield, Dublin 4, Ireland, D04 V1W8}
\author{Samuel Valach}
\affiliation{Faculty of Mathematics and Physics, University of Ljubljana, SI-1000, Ljubljana, Slovenia}
 
\begin{abstract}
At large $N$, the analytic structure of holographic thermal correlation functions
contains non-trivial information about the near-singularity region of dual black
holes. In particular, complex-time singularities can be related to geodesics that
probe the black hole interior and ``bounce off'' the singularity. We analyse the
asymptotic behaviour of holographic OPE data entering scalar two-point functions
in states dual to black branes whose blackening factors truncate at finite order
in inverse powers of the radial coordinate. For this class of geometries, we find
a direct relation between the power-law exponent of these bouncing singularities
and the Kasner exponent associated with the time direction near the black hole
singularity. We discuss the regime of validity of this relation and comment on
possible extensions to the transverse spatial directions.

\end{abstract}

\maketitle


\section{Introduction and summary}
\label{sec:introduction}

The AdS/CFT correspondence \cite{Maldacena:1997re, Gubser:1998bc, Witten:1998qj} provides a powerful framework for probing the deep interior of black holes using field-theoretical methods. 
For example, the analytic structure of large-$N$ correlation functions at finite temperature $T=1/\beta$ in the boundary CFT encodes signatures of the bulk curvature singularity \cite{Fidkowski:2003nf,Festuccia:2005pi}. Specifically, when analytically continued to complex times, boundary retarded propagators exhibit branch-point singularities -- often referred to as \textit{bouncing singularities} -- whose locations, $t_c$, correspond to the boundary time difference between the endpoints of geodesics that ``bounce off'' the black hole singularity \cite{Parisini:2023nbd,Ceplak:2024bja,Grozdanov:2026cut}. In the language of Wightman functions, $G(t,x\equiv\abs{\vec{x}})=\expval{\mathcal{O}(t,\vec{x})\mathcal{O}(0,0)}_\beta$, of a scalar field $\mathcal{O}$ with conformal dimension $\Delta$, these are the singularities of the ``current sector'' (composed of all integer-dimensional operator exchanges) of the form
\begin{equation}\label{e.bs}
    G_T(t)\equiv G_T(t,x=0) \sim \frac{1}{(t_c - t)^{2\Delta - b_t}},
\end{equation}
and can be effectively extracted using various OPE, WKB and numerical techniques \cite{Parisini:2023nbd,Ceplak:2024bja,Buric:2025fye,Valach:2025saf,Barrat:2025twb,Afkhami-Jeddi:2025wra,Dodelson:2025jff,Ceplak:2025dds,Jia:2025jbi,AliAhmad:2026wem,Jia:2026pmv,Araya:2026shz,Giombi:2026kdz,Grozdanov:2026cut,Jia:2026ryl,Arnaudo:2026tcy,Grozdanov:2026ktq,Buric:2026qsp,Buric:2026cmn,Arnaudo:2026axe,Barrat:2026jfg,Grozdanov:2026lnc}.

While previous studies have shown that the location of these singularities in the complex-time plane encodes information about the black hole curvature singularity, the physical information carried by the exponent $b_t$ has remained elusive. This brings us to the main question we want to answer: Imagine you are given a CFT with a correlator $G_T(t)$ satisfying \eqref{e.bs}. Can the value of the exponent $b_t$ reveal something about the near-singularity behaviour of the dual geometry? The answer turns out to be yes.

Near a generic spacelike singularity, the spacetime is typically described by a Kasner model \cite{Kasner:1921zz,Belinsky:1970ew}
\begin{align}
\label{eq:Kasner}
    \dd s^2=-\dd\sigma^2+C_t\,\sigma^{2p_t}\dd t^2 + \sum_{i=1}^{d-1}C_i\,\sigma^{2p_i}\dd x_i^2,
\end{align}
characterised by a set of Kasner scaling exponents $p_t$ and $p_i$,  where $C_t$ and $C_i$ are some constants.
The main result of our paper is that, in simple black brane models, defined in more detail in Section~\ref{sec:Kasner}, there exists an explicit relation between the time-scaling Kasner exponent and the power $b_t$ of the bouncing singularity
\begin{align}\label{e.maincoon}
    b_t = \frac{d-1}{1-p_t}\,,\qquad \Longleftrightarrow\qquad   p_t = \frac{1+b_t-d}{b_t}\,.
\end{align}
We conjecture that this relation holds more generally for a broad class of black brane solutions described by metrics of the form \eqref{eq:BBmet}, and for correlation functions at vanishing spatial separation.
The encoding of Kasner exponents directly into the power of the bouncing singularity offers an effective avenue for probing near-singularity geometries via the dual CFT.

\section{Gravitational Setup}
\label{sec:Kasner}

Consider a gravitational theory in $(d+1)$-dimensional anti-de Sitter spacetime that contains a planar black hole solution of the form
\begin{align}
\label{eq:BBmet}
    \dd s^2 = - r^2 f(r) \dd t^2 + \frac{\dd r^2}{r^2 f(r)}+ r^2\dd\vec{x}_{d-1}^{\,2}\,,
\end{align}
where $r$ denotes the radial direction with the black hole singularity located at the origin and the AdS boundary at $r\to \infty$.
We will limit ourselves to spacetimes with even $d$ and assume that the blackening factor can be written as an expansion in inverse powers of the radius, truncating at order $K$
\begin{equation}\label{eq:blackening}
    f(r) = 1 - \sum_{j = 1}^{K} \frac{\mu_j}{r^{2j}}\,,
\end{equation}
and we label the solutions by a pair of integers $(d,K)$. 
We assume that $\mu_K >0$, which ensures that the singularity at $r=0$ is spacelike. 
We remain agnostic about the origin of these solutions.%
\footnote{They can arise, for example, in charged black brane solutions \cite{Chamblin:1999hg, Chamblin:1999tk, Brecher:2004gn}, in gravitational theories with stringy corrections truncated to some order \cite{Gubser:1998nz,Pawelczyk:1998pb}, or in truncated models of Gauss-Bonnet gravity \cite{Boulware:1985wk, Cai:2001dz} and quasi-topological gravity \cite{Myers:2010jv}.}
Indeed, some of them may not correspond to physically sensible solutions.
For our purposes, however, they serve as effective toy models for analysing holographic correlation functions.

Near the singularity at $r \to 0$, the most singular term of the blackening factor dominates, $f(r) \sim -\frac{\mu_K}{r^{2K}}$, 
and the metric becomes
\begin{align}
    \dd s^2=\mu_K r^{2-2K}\dd t^2-\frac{r^{2K-2}}{\mu_K}\dd r^2+r^2\dd\vec{x}_{d-1}^{\,2}\,.
\end{align}
Using the coordinate transformation $\dd\sigma= \frac{r^{K-1}\dd r}{\sqrt{\mu_K}}$, where $\sigma\to0$ corresponds to the near-singularity regime, this metric can be rewritten in the Kasner form~\eqref{eq:Kasner}
with the exponents
\begin{align}
    \label{eq:KasExp}
    p_t = \frac{1-K}{K}\,,\qquad p_i = \frac{1}{K}\,.
\end{align}
These exponents satisfy the vacuum Kasner constraints of pure gravity $\sum_a p_a = \sum_a p_a^2 = 1$ only for $K = d/2$, that is for the Schwarzschild--AdS$_{d+1}$ black brane solution. For other values of $K$, the geometry must be supported by additional matter fields or arise from higher-derivative corrections, and the constraints are violated.


\section{Bouncing Singularity}
\label{sec:CFT}

In this section, we compute holographic thermal two-point functions of identical scalar primary operators $\mathcal{O}_{\Delta}$ of conformal dimension $\Delta$, dual to minimally coupled bulk scalar fields, in the black hole backgrounds defined in \eqref{eq:BBmet}.
At vanishing spatial separation, $\vec{x}=0$, the relevant OPE sector of these correlators develops a singularity at $t_c$, of the form given in \eqref{e.bs} with the corresponding exponent depending on $\Delta$ and on a parameter $b_t$.
By scanning the space of $(d,K)$ pairs, we find that, for this class of geometries, $b_t$ is given by a simple function of $d$ and $K$, and therefore directly encodes information about the Kasner exponent $p_t$.

\subsection{OPE analysis}

We follow the method introduced in \cite{Fitzpatrick:2019zqz} to compute the coefficients $\Lambda_n$ associated with the current sector%
\footnote{The method determines the coefficients of those terms in the small-$\tau$ expansion of the correlator whose exponents are of the form  $-2\Delta+ k$, with generic $\Delta$ and integer $k$. In physically sensible examples, these include exchanges of the stress tensor, conserved currents, and their composites. We therefore refer to this contribution as the current sector.}
in the short-time expansion of the thermal correlator,
\begin{equation}\label{e.modelka}
    G_T(\tau)=\frac{1}{\tau^{2\Delta}}\sum_{n=0}^\infty\Lambda_n(\mu_j)\,\tau^{2 n},
\end{equation}
where $\tau=it$.
The expansion can be constructed for any integer $K$ and even $d\geq 2$; for simplicity, we take $\Delta$ to be real and non-integer.
Details are given in Appendix~\ref{a.recurr}.

As shown in \cite{Parisini:2023nbd,Ceplak:2024bja}, the radius of convergence of the expansion \eqref{e.modelka} is controlled by the bouncing singularity at $\tau=\tau_c$, near which the current sector behaves as
\begin{equation}\label{e.bss}
    G_T(\tau) \sim \frac{1}{(\tau_c - \tau)^{2\Delta - b_t}}\,.
\end{equation}
The parameters $\tau_c= i\,t_c$ and $b_t$ are then fixed by the large-$n$ behaviour of the coefficients $\Lambda_n(\mu_j)$.

Our aim is to determine how the exponent $b_t$ depends on $d$ and $K$.
To this end, we compute the expansion \eqref{e.modelka} for a range of backgrounds labelled by pairs $(d,K)$.
As we show in Section~\ref{ss.ssd}, it is sufficient, and computationally much less expensive, to turn on only the coefficient $\mu_K\equiv\mu$, setting $\mu_j=0$ for all $j\neq K$.
This gives
\begin{equation}
\label{eq:BFSimple}
    f(r)=1-\frac{\mu}{r^{2K}}\,,
\end{equation}
from which it follows that the only non-zero coefficients in the expansion
\eqref{e.modelka} multiply $\tau^{2Kn}$, which we relabel as
$\Lambda_n\,\tau^{2Kn}$ in what follows.

The leading large-$n$ behaviour of the expansion coefficients is found to be \cite{Ceplak:2024bja}
\begin{align}\label{e.lblb}
    \Lambda_n \sim \left[\mu\,a(d,K)\right]^n\,n^{2\Delta-b_t(d,K)-1},
\end{align}
where we have suppressed the overall constant prefactor, as well as terms that are subleading at large $n$.
Near the radius of convergence of \eqref{e.modelka}, these large-$n$ coefficients dominate and can be resummed to produce the singular behaviour \eqref{e.bss}.
The exponent $b_t$ can hence be read off from the asymptotic behaviour of the holographic data.

The position of the singularity is fixed by the exponential growth of the coefficients,
\begin{equation}
    \tau_c^{2K}=\left[\mu\,a(d,K)\right]^{-1}.
\end{equation}
This equation has $2K$ complex solutions for $\tau_c$.
Among these solutions, a complex-conjugate pair that has real part equal to $\beta/2$ corresponds to the separation between boundary points in the two-sided black hole geometry that are connected by a bouncing geodesic \cite{Fidkowski:2003nf}.
However, since this location depends on the temperature $\beta$, which is in general a function of all the coefficients $\mu_j$ in the blackening factor \eqref{eq:blackening}, it is not clear how much detailed information about the near-singularity behaviour of the metric, such as the value of $K$, can be extracted from it.
We do not focus on this observable in this paper.

\begin{figure}[t]
    \centering
    \includegraphics[width=0.92\columnwidth]{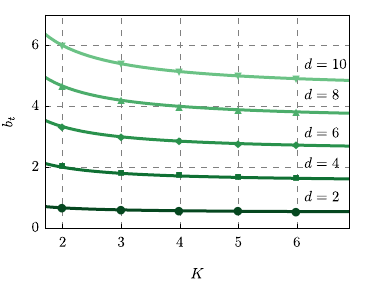}
    \vspace{-0.3cm}
    \caption{Comparison of the numerical data presented in Table~\ref{t.tab1} with the analytic relation between $(d,K)$ and the exponent $b_t$, for several spacetime dimensions. The data are computed for $\Delta=3/2$ using coefficients up to $n=400$.}
    \label{fig:dataplot}
\end{figure}

The exponent $b_t$, on the other hand, is independent of $\mu$, and the entire dependence on the conformal dimension sits in the term $2\Delta$ in the exponent of \eqref{e.lblb}, as we have verified by repeating the fits at several values of $\Delta$. Taking $\Delta=3/2$ as a representative value, we collect several
examples in Table~\ref{t.tab1} and Figure~\ref{fig:dataplot}.
These values are obtained by computing the asymptotic behaviour of the coefficients $\Lambda_n$ and performing a numerical fit to the asymptotic form in \eqref{e.lblb}.
\begin{table}
\setlength{\tabcolsep}{8pt}
\begin{tabular}{|c|c||S||r@{${}={}$}S|}
\hline
$d$ & $K$ & {\textcolor{dred}{$b_t$ (numeric)}} & \multicolumn{2}{c|}{$b_t$ (analytic)} \\
\hline\hline
2 & 2 & \textcolor{dred}{0.66669} & $2/3$ & 0.66667 \\ \hline
2 & 3 & \textcolor{dred}{0.60003} & $3/5$ & 0.6 \\ \hline
2 & 4 & \textcolor{dred}{0.57147} & $4/7$ & 0.57143 \\ \hline
2 & 5 & \textcolor{dred}{0.55559} & $5/9$ & 0.55556 \\ \hline
2 & 6 & \textcolor{dred}{0.54549} & $6/11$ & 0.54546 \\ \hline\hline
4 & 2 & \textcolor{dred}{1.99992} & \multicolumn{2}{c|}{$2$} \\ \hline
4 & 3 & \textcolor{dred}{1.80004} & $9/5$ & 1.8 \\ \hline
4 & 4 & \textcolor{dred}{1.71431} & $12/7$ & 1.71429 \\ \hline
4 & 5 & \textcolor{dred}{1.66671} & $5/3$ & 1.66667 \\ \hline
4 & 6 & \textcolor{dred}{1.63639} & $18/11$ & 1.63636 \\ \hline\hline
6 & 2 & \textcolor{dred}{3.33337} & $10/3$ & 3.33333 \\ \hline
6 & 3 & \textcolor{dred}{2.99993} & \multicolumn{2}{c|}{$3$} \\ \hline
6 & 4 & \textcolor{dred}{2.85717} & $20/7$ & 2.85714 \\ \hline
6 & 5 & \textcolor{dred}{2.77784} & $25/9$ & 2.77778 \\ \hline
6 & 6 & \textcolor{dred}{2.72732} & $30/11$ & 2.72727 \\ \hline\hline
8 & 2 & \textcolor{dred}{4.66611} & $14/3$ & 4.66667 \\ \hline
8 & 3 & \textcolor{dred}{4.20037} & $21/5$ & 4.2 \\ \hline
8 & 4 & \textcolor{dred}{3.99989} & \multicolumn{2}{c|}{$4$} \\ \hline
8 & 5 & \textcolor{dred}{3.88889} & $35/9$ & 3.88889 \\ \hline
8 & 6 & \textcolor{dred}{3.81819} & $42/11$ & 3.81818 \\ \hline\hline
10 & 2 & \textcolor{dred}{5.99998} & \multicolumn{2}{c|}{$6$} \\ \hline
10 & 3 & \textcolor{dred}{5.39997} & $27/5$ & 5.4 \\ \hline
10 & 4 & \textcolor{dred}{5.14473} & $36/7$ & 5.14286 \\ \hline
10 & 5 & \textcolor{dred}{4.99983} & \multicolumn{2}{c|}{$5$} \\ \hline
10 & 6 & \textcolor{dred}{4.90898} & $54/11$ & 4.90909 \\ \hline
\end{tabular}
\caption{Numerical values of $b_t(d,K)$ for various choices of $d$ and $K$. 
The third column is obtained by fitting the numerical data for $\Lambda_n$, for a scalar with $\Delta=3/2$, to the asymptotic form in \eqref{e.lblb}. 
The quoted accuracy is based on 400 data points. 
The final column gives the analytic value obtained from \eqref{eq:Proposal}. 
We find good agreement between the numerical and analytic results.}
\label{t.tab1}
\end{table}
We observe that the numerical values are in good agreement with 
\begin{align}\label{eq:Proposal}
    b_t = \frac{K(d-1)}{2K-1}\,,
\end{align}
for $K>1$ and $d>1$ that is even. We comment on the case $K=1$ in Section~\ref{s.ext}.
Combining this expression with \eqref{eq:KasExp}, we obtain a relation between the Kasner exponent $p_t$ and the power of the bouncing singularity,
\begin{align}\label{e.submaincoon}
    b_t = \frac{d-1}{1-p_t}\,,\qquad \Longleftrightarrow\qquad   p_t = \frac{1+b_t-d}{b_t}\,.
\end{align}
In other words, the strength of the divergence at the first non-trivial singularity in the complex-time plane of the holographic correlator is directly related to the near-singularity behaviour of the black hole metric.

\subsection{Approach to the exponent}\label{ss.ssd}

\begin{figure*}[ht!]
    \centering
    \includegraphics[width=0.90\textwidth]
        {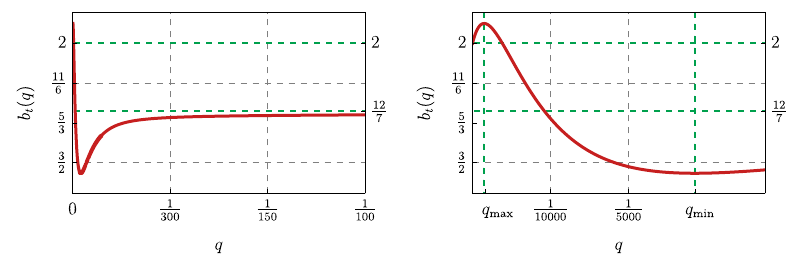}
    \caption{Behaviour of $b_t$ as a function of $q$ for the background \eqref{e.deftoy} with $\mu=1$. We fix $\Delta=3/2$ and fit the first $m=220$ OPE data $\Lambda_n$. The function $b_t(q)$ quickly converges to the saturation value $12/7$ (left plot). Zooming in to the region of sufficiently small $q$ (right plot), we see the local minimum and maximum that describe the non-trivial transition window.}
    \label{f.toyfixm}
\end{figure*}

The result \eqref{e.submaincoon} assumes that the blackening factor takes the form \eqref{eq:BFSimple}, with all other $\mu_j$ being set to 0.
This leads to the exponent $b_t$ being independent of the parameter $\mu$. 
On the other hand, if several $\mu_j$ are non-vanishing, one can construct non-trivial dimensionless parameters on which $b_t$ can in principle depend. 
In this section, we show that as $n$ is increased, terms in \eqref{eq:blackening} that are subleading in the near-singularity limit are irrelevant.
Let us consider $d=4$ and
\begin{equation}\label{e.deftoy}
    f(r)=1-\frac{\mu}{r^4}-\frac{q}{r^8}\,.
\end{equation}
We can rescale the radial coordinate to effectively set $\mu=1$, which leaves $q$ as a free dimensionless parameter.
From our previous analysis, one expects
\begin{equation}\label{e.desire}
    K=\begin{cases}
       2,&\,q=0\\
       4,&\,q>0\,,
    \end{cases}\qq{thus}\!b_t=\begin{cases}
       2,&\,q=0\\
       12/7,&\,q>0\,.
    \end{cases}
\end{equation}
Let us analyse this transition.
In Figure~\ref{f.toyfixm}, we plot the behaviour of $b_t$ as a function of $q$ in the \eqref{e.deftoy} background. 
This value of $b_t$ is obtained by numerically calculating the first $m=220$ $\Lambda_n$ and analysing their behaviour as $n$ is increased.
The function $b_t(q)$ for a fixed $m$ converges towards $b_t=12/7$ as $q$ is increased. However, for sufficiently small $q$ there is a transition window where the function transitions between the expected values \eqref{e.desire}.
We observe a
local minimum and maximum at $q_{\rm min}$, respectively $q_{\rm max}$. The values of $b_t(q)$ at these local extrema are $\Delta$-independent and do not change significantly with rising $m$. We can therefore use the locations of $q_{\rm min}$ and $q_{\rm max}$ as an approximate measure of the size of the transition window as a function of $m$. We find that for large enough $m$
\begin{equation}
    q_{\rm min,\,max}\sim m^{-
    \nu}, \qquad \nu\approx1.35\,.
\end{equation}
Hence, as $m$ is increased, this transition region shrinks, and in the limit
$m\to\infty$, which is the regime responsible for the leading singular behaviour
of the correlation function \eqref{e.bss}, it vanishes, leaving us with an
effectively discontinuous behaviour in $q$, as described in \eqref{e.desire}.

We expect the same qualitative behaviour for generic blackening factors of the type \eqref{eq:blackening}, with several non-vanishing $\mu_j$. 
The observed function $b_t$, given in \eqref{eq:Proposal}, is monotonically decreasing in $K$ for all $K>1/2$ and all $d>1$. 
Hence, for any finite $m$, one can imagine a cascade of transitions to $b_t$ values associated with terms which become dominant as one approaches the singularity. 
As $m$ is increased, $b_t$ associated with the most singular contribution to $f(r)$ near the singularity will become dominant for smaller values of the parameters. Extrapolating to infinite $m$, we conclude that the $b_t$ exponent observed in holographic correlation functions is uniquely fixed by the leading power $K$ and dimension $d$.


\section{Extensions and Limitations}
\label{s.ext}

For $K=1$, or when allowing non-zero spatial separation, one cannot relate the Kasner exponents to the exponent $b_t$. 
This is consistent with expectations, since in these cases the relevant bouncing geodesic does not exist. 
The first non-trivial singularity of the boundary correlator is then not associated with a probe of the near-singularity region.

\subsection{Vanishing temporal separation}\label{ss.notime}

Similarly to the calculation performed in Section~\ref{sec:CFT}, one can extract the data for the situation where we set $\tau=0$, in which case the OPE decomposition reads \cite{Araya:2026shz}
\begin{equation}
    G_T(x)=\frac{1}{x^{2\Delta}}\sum_{n=0}^\infty\Lambda_n^{(x)}\,x^{2Kn}\,,
\end{equation}
with the OPE data $\Lambda_n^{(x)}$ admitting the same asymptotic form \eqref{e.lblb}. Following the same approach, we find (see Table~\ref{t.tab2} for a summary)
\begin{equation}\label{e.xxxfifa}
    b_x=\frac{d-1}{2}\,.
\end{equation}
\begin{table}
\centering
\settowidth{\lostwd}{\,\,\,\,\,0.5000\,\,\,\,\,}%
\addtolength{\lostwd}{2\tabcolsep}%
\addtolength{\lostwd}{\arrayrulewidth}%
\setlength{\tabcolsep}{\dimexpr\tabcolsep+\lostwd/8\relax}%
\begin{tabular}{|c|c||c||c|}
\hline
d & K & \!\!\color{dred}{$b_x$ \!(numeric)}\!\! & \!\!$b_x$ \!(analytic)\!\! \\
\hline\hline
\,\,2\,\, & \,\,2\,\, & \,\,\,\,\,\color{dred}{0.4964}\,\,\,\,\, & \,\,\,\,\,1/2\,\,\,\,\, \\ \hline
2 & 3 & \color{dred}{0.4992} & 1/2 \\ \hline
2 & 4 & \color{dred}{0.5005} & 1/2 \\ \hline
2 & 5 & \color{dred}{0.5104} & 1/2 \\ \hline\hline
4 & 2 & \color{dred}{1.4966} & 3/2 \\ \hline
4 & 3 & \color{dred}{1.5001} & 3/2 \\ \hline
4 & 4 & \color{dred}{1.4981} & 3/2 \\ \hline
4 & 5 & \color{dred}{1.5045} & 3/2 \\ \hline\hline
6 & 2 & \color{dred}{2.4929} & 5/2 \\ \hline
6 & 3 & \color{dred}{2.4980} & 5/2 \\ \hline
6 & 4 & \color{dred}{2.5003} & 5/2 \\ \hline
6 & 5 & \color{dred}{2.4992} & 5/2 \\ \hline\hline
8 & 2 & \color{dred}{3.4989} & 7/2 \\ \hline
8 & 3 & \color{dred}{3.5011} & 7/2 \\ \hline
8 & 4 & \color{dred}{3.5019} & 7/2 \\ \hline
8 & 5 & \color{dred}{3.4983} & 7/2 \\ \hline\hline
10 & 2 & \color{dred}{4.4749} & 9/2 \\ \hline
10 & 3 & \color{dred}{4.5066} & 9/2 \\ \hline
10 & 4 & \color{dred}{4.5090} & 9/2 \\ \hline
\end{tabular}
\caption{Numerical values of $b_x (d,K)$ obtained from asymptotic fitting compared to  $(d-1)/2$. The data is for $\Delta =3/2$ with up to $n=400$ data points.}
\label{t.tab2}
\end{table}
The lack of the $K$-dependence is related to the fact that the geodesic structure is sharply different for the setups with boundary anchoring $(t,x=0)$ and those with $(t,x\neq0)$.  For $x\neq0$, the first non-trivial singularity of $G_T(\tau,x)$ corresponds to a complex bulk-cone geodesic that does not probe the black hole interior \cite{Araya:2026shz,Buric:2026cmn}. It is thus natural to expect that any setup with non-zero $x$ and arbitrary $t$ leads to an exponent $b_x(d,K)$ which is insensitive to the region near the curvature singularity and does not contain any information about the near-singularity Kasner exponents.
Interestingly, $b_x$ is exactly the formal $K\to\infty$ limit of the expression for $b_t$, or equivalently the limit $p_t\to -1$, in which all $p_i\to0$ and the transverse directions stop scaling with the near-singularity coordinate $\sigma$.

\subsection{Mixed momentum space}\label{ss.momsntumm}

Consider now a setup in which one keeps $t\neq0$ and Fourier transforms the remaining coordinates in the boundary CFT -- this introduces the notion of the mixed momentum space, where the description is in terms of $(t,k)$, where $k\equiv|\vec{k}|$ and $\vec{k}$ is the spatial momentum. In \cite{Dodelson:2025jff} it was shown that, starting with the thermal product formula \cite{Dodelson:2023vrw}, one is able to recover the structure of the bouncing singularities in the mixed momentum space. Performing a large-frequency expansion and a direct Fourier transform $\omega\to t$ of the formal transseries, the authors derived the locations of the bouncing singularities in $(t,k)$ space, matching the known results from the $(t,x=0)$ space, as well as the universal power of these singularities. This power, which we denote $b_k$, turns out to be completely independent of the blackening factor and, similarly to the $(t=0,x)$ case, depends only on the dimension:
\begin{equation}\label{e.bpeq}
    b_k(d)=d-1\,.
\end{equation}
Importantly, $b_k$ differs from both $b_t$ and $b_x$. This is due to a non-trivial relation between $(t,k)$ and $(t,x)$ spaces. Indeed, the additional Fourier-transform integral not only affects the power of the singularities but in some cases also removes singularities from the complex plane, see \cite{Grozdanov:2026cut} for more details.

In summary, comparing Equations~\eqref{eq:Proposal} and \eqref{e.bpeq}, we see that the Fourier transformation from mixed momentum space to position space reorganises the branch-point singularity structure. In particular, the momentum integral changes the power of the singularity and exposes a $K$-dependence that is not visible in the $b_k$ exponent.

\subsection{\texorpdfstring{$K=1$}{K=1}}\label{ss.Kis1}

Computing $b_t$ or $b_x$, we notice that for the cases where $K=1$ (e.g., BTZ black hole in $d=2$ or self-dual axion model in $d=3$ \cite{Andrade:2013gsa}), one is unable to perform the fit to \eqref{e.lblb} as discussed above and the procedure for obtaining the relation between $b_t$ and the Kasner exponents fails.

This is to be expected since, for $K=1$, one has $r^2 f(r) = r^2 - \mu$. Consequently, the time component of the metric, $g_{tt} = -r^2 f(r)$, approaches a constant rather than expanding as $r \to 0$, hence the assumptions of the existence theorem \cite{Grozdanov:2026cut} for bouncing singularities are not satisfied. In other words, the first non-trivial singularity of $G_T(t)$ is not associated with a bouncing geodesic and should thus be insensitive to the physics of the deep black hole interior.


\section{Discussion}\label{s.discs}

We have shown that, for the class of black hole solutions described by the
blackening factor \eqref{eq:blackening}, the exponent $b_t$ governing the
bouncing singularity of the correlator $G_T(t)$ encodes the behaviour of the
metric near the black hole singularity. We used the relation \eqref{eq:Proposal}
to connect $b_t$ to the temporal Kasner exponent $p_t$ in \eqref{e.maincoon}.

The transverse exponent $p_i$ also depends non-trivially on $K$, but it does not
appear in this relation: the singularity \eqref{e.bs} arises from bouncing
geodesics that are purely radial and therefore do not probe the transverse
directions. Disentangling the roles of $p_t$ and $p_i$ therefore requires
geometries in which the blackening factors entering the time and radial
components of the metric have different near-singularity behaviour, while the
transverse directions remain isotropic. We initiate this analysis in
Appendix~\ref{a.unequalblack}; however, at our current computational accuracy,
the extracted values of $b_t$ do not converge to a simple analytic expression.

While the expression relating $b_t$ to $d$ and $K$ is specific to the family of
solutions analysed here, we expect the relation \eqref{e.maincoon} between the holographic exponent
and the temporal Kasner exponent $p_t$ to hold more broadly for spacetimes described by \eqref{eq:BBmet}. We have focused on
even $d$ because of technical subtleties in the near-boundary OPE techniques,
but we see no physical obstruction to the relation holding for arbitrary $d$.%
\footnote{Momentum-space methods
\cite{Dodelson:2025jff,Hartnoll:2026vhu,Xiao:2026pir} work in arbitrary $d$,
although potential subtleties in odd $d$ were highlighted in
\cite{Serantes:2022mgl}.}
It would be useful to verify this explicitly.

It would be very interesting to analyse black hole solutions in which the Kasner exponents are not rational, such as those arising in theories with relevant deformations \cite{Frenkel:2020ysx,Hartnoll:2020rwq}, in holographic superconductors \cite{Hartnoll:2020fhc,Sword:2021pfm}, and in hairy black holes more generally \cite{Cai:2020wrp,Mansoori:2021wxf,Arean:2024pzo}.

The relation \eqref{eq:Proposal} was established numerically by fitting the asymptotics of the OPE data across a range of $(d,K)$.
An analytic derivation would clarify whether the truncation of the blackening factor is essential, and whether the relation survives in geometries whose near-singularity behaviour is not governed by a simple power.

We have worked entirely in position space, where an analytic derivation is hard to obtain directly. Recent works
\cite{Hartnoll:2026vhu,Xiao:2026pir}%
\footnote{See also \cite{Coviello:2026tol} for a similar analysis in quantum black holes, and \cite{Li:2026gqf} for a related discussion involving other observables.}
address related questions in momentum space, showing that Kasner exponents can be extracted from the analytic behaviour of quasinormal modes at large overtone number. In particular, this gives access to the transverse exponent $p_i$, which our analysis does not probe.

As shown in Figure~\ref{f.toyfixm}, the strict $n\to\infty$ behaviour of the
expansion coefficients $\Lambda_n$ is sensitive only to the leading behaviour of
the metric near the singularity.
At finite $n$, however, there is a transition regime in which the coefficients
interpolate between the effects of the different terms in the blackening factor
that diverge near the singularity.
We have demonstrated this only for a solution with two non-trivial terms in the
blackening factor, but we see no obstruction to this behaviour generalising to
blackening factors with more terms.

This raises the interesting possibility that this transition takes the form of
a ``cascade'': for fixed values of the parameters $\mu_j$, an
effective exponent extracted from finite-$n$ data detects the individual terms in
$f(r)$ in succession before settling, at sufficiently large $n$, to the asymptotic value
controlled by the most singular term. 
This mechanism is reminiscent of the transition observed in the large-overtone
behaviour of quasinormal modes of black holes whose interiors contain different
Kasner epochs \cite{Hartnoll:2026vhu}.
It would be interesting to determine whether an analogous transition is visible
in the finite-$n$ approach to the exponent $b_t$.

Looking further ahead, \cite{Hartnoll:2026vhu,Xiao:2026pir} have also discussed a possible route to observing the full Belinski--Khalatnikov--Lifshitz--Mixmaster dynamics \cite{Belinsky:1970ew,Misner:1969hg,DeClerck:2023fax,Caceres:2026mug} in the asymptotic structure of quasinormal modes.
It would be interesting to find the analogue of these results in the position-space approach presented here and establish a direct dictionary between those momentum-space results and the findings of this paper, perhaps using the framework developed in \cite{Arnaudo:2026axe,Barrat:2026jfg}.


\section*{Acknowledgements}

We would like to thank Chantelle Esper, Sašo Grozdanov, Sean Hartnoll, and Andrei Parnachev for discussions. This work  is supported in part by Taighde \'Eireann - Research Ireland under the grant agreement 22/EPSRC/3832. S.V.\ is supported by the Marie Skłodowska-Curie Actions programme GA-101177446 and the Slovenian Research and Innovation Agency (ARIS), contract number 5110-18/2025-5.


\appendix
\section{Generalised recurrence relations}\label{a.recurr}

In this appendix we describe the generalisation of the recurrence relations method first discussed in \cite{Buric:2025fye,Ceplak:2025dds} as an effective technical implementation of the near-boundary expansion \cite{Fitzpatrick:2019zqz} (see also \cite{Li:2019tpf,Karlsson:2022osn,Huang:2022vet,Esper:2023jeq} for its generalisations). This method allows one to compute the current sector OPE data for any theory dual to a $(d+1)$-dimensional black brane
\begin{align}
    \dd s^2&=r^2f(r)\,\dd\tau^2+\frac{\dd r^2}{r^2h(r)}+r^2\dd\vec{x}^{\phantom{.}2},\\
    f(r)&=P_{K}\left(\tfrac{1}{r^2}\right),\,\, h(r)=P_{L}\left(\tfrac{1}{r^2}\right),
\end{align}
where $P_{K}\left(\frac{1}{r^2}\right)$ and $P_{L}\left(\frac{1}{r^2}\right)$ denote arbitrary polynomials in $r^{-2}$ of order $K$, resp.~$L$. The current sector of the dual correlator can then be expressed as
\begin{equation}\label{e.genelegante}
    G_T(\tau,x)=\frac{1}{(\tau^2+x^2)^\Delta}\,\sum_{i=0}^\infty\,\sum_{j=0}^{\lfloor\frac{i}{2}\rfloor}\,a_{j,i-j}^i\,\frac{x^{2j}(\tau^2+x^2)^{(i-j)}}{r^{2\,i}},
\end{equation}
where $x\equiv\abs{\vec{x}}$ and the coefficients $a^i_{j,k}$ are obtained by the following algorithm:

Start by setting
    \begin{equation}\label{e.prvkaprea}
    a^i_{j,k}=
    \begin{cases}
    1&\text{for}\quad(i,j,k)=(0,0,0)  \\
    0&\forall\,(i,j,k)\notin I\cup\{(0,0,0)\},
    \end{cases}
    \end{equation}
where the index set $I=\{\,(i,j,k)\mid i\geq1\wedge0\leq j\leq\lfloor\frac{i}{2}\rfloor\wedge-\lceil\frac{i}{2}\rceil\leq k\leq i-j\}$. To determine the remaining values of $a^i_{j,k}$ use the following steps:
    \begin{itemize}
        \item For each $i\geq1$,
        \item For each $k\in\left[-\lceil\frac{i}{2}\rceil,\,i\right]$,
        \item For each $j\in\left[0,\,\lfloor\frac{i}{2}\rfloor\right]$,
        \item Compute $a^i_{j,k}$ using
    \begin{equation}\label{e.mmacinka}
            a^i_{j,k}=-\frac{1}{C_{0,1,1}^{\,i,j,k}}\sum_{(a,b,c)\in J\backslash\{(0,0,0)\}}a^{i-a}_{j-b,\,k-c}\cdot C^{\,i-a,\,j-b,\,k-c}_{a,\,b+1,\,c+1},
    \end{equation}
    \end{itemize}
where the index set $J=\{(a,b,c)\mid a\in[0,K+L]\wedge b\in[-1,1]\wedge c\in[-1,1]\}$ and the $C_{a,b,c}^{\,i,j,k}$ is a $(K+L+1)\times3\times3$ dimensional tensor for any fixed $i,\,j,\,k$, defined implicitly by the following relation, which expresses the action of the Klein-Gordon operator on the ansatz \cite{Fitzpatrick:2019zqz} (see also \cite{Buric:2025fye})
{\small{
\begin{align}
    \sum_{a=0}^{K+L}&\sum_{b=0}^2\sum_{c=0}^2\,C_{a,b,c}^{\,i,j,k}\,\,\frac{\rho^{2b}w^{2c}}{r^{2a}}\equiv\\ &A_1+A_2\,f(r)+A_{3}\,f(r)h(r)+A_4\,\dv{}{r}\left(f(r)h(r)\right),\nonumber
\end{align}}}%
where $A_1$, $A_2$, $A_3$ and $A_4$ are functions of $\rho$ and $w$ given explicitly by
{\allowdisplaybreaks
{\small{
\begin{align}
    A_1=&-4 \rho ^2 (k-\Delta ) \big(-2 (\Delta +1) \left(\rho ^2+1\right)\nonumber\\
    &\hspace{1.5cm}+2 k \left(\rho ^2-w^2+1\right)+(2 \Delta +1) w^2\big),\\
    A_2=&2 \rho ^2 w^2 \big(2 k (d+4 j-1)+\Delta  \big(d \left(w^2-2\right)\nonumber\\
    &-8 j-\Delta  w^2+2\big)\big)+4 j w^4 (d+2 j-3)\nonumber\\
    &\hspace{2.3cm}+8 \rho ^4 (-\Delta +k-1) (k-\Delta ),\\
    A_3=&2 \rho ^2 \big(2 w^4 (-i+j+k) (d+2 (-i+j+k))\nonumber\\
    &-2 k w^2 (d-4 i+4 j+4 k-2)-\Delta  \big(w^4 (d+4 (-i\nonumber\\
    &+j+k))-2 w^2 (d-4 i+4 j+6 k-2)+8 k-4\big)\nonumber\\
    &\hspace{2.05cm}+4 (k-1) k+\Delta ^2
   \left(w^2-2\right)^2\big),\\
    A_4=&\rho ^2 r w^2 \left(2 \Delta -\left(w^2 (\Delta +2 i-2 j)\right)+2 k \left(w^2-1\right)\right).
\end{align}}}}%

Let us finish by focusing on the concrete cases examined in this paper. Assume $K=L$ and the blackening factor \eqref{eq:blackening}. Setting $x=0$, the only non-zero terms in the current sector of the correlator \eqref{e.genelegante} are
\begin{equation}
    G_{T}(\tau)=\frac{1}{\tau^{2\Delta}}\sum_{n=0}^\infty\Lambda_n\,\mu^n\,\tau^{2Kn},\,\,\,\Lambda_n\equiv\frac{a_{0,Kn}^{Kn}}{\mu^n}.
\end{equation}
If one is instead interested in $\tau=0$, equation \eqref{e.genelegante} immediately yields
\begin{equation}
    G_{T}(x)=\frac{1}{x^{2\Delta}}\sum_{n=0}^\infty\Lambda^{(x)}_n\,\mu^n\,x^{2Kn},\,\,\,\Lambda^{(x)}_n\!\equiv\!\frac{1}{\mu^n}\!\sum_{j=0}^{\lfloor\frac{Kn}{2}\rfloor}a_{j,Kn-j}^{Kn}.
\end{equation}
This way one can effectively compute $\Lambda_n$ and $\Lambda_n^{(x)}$ to orders $n\sim300-1000$ on a standard laptop.

\begin{table}[!t]
\centering
\begin{tabular}{|c|c|c|c||c|c|c|c||c|c|c|c||c|c|c|c|}
\hline
d & K & L & \color{dred}{\,\,\,\,\,$b_t$\,\,\,\,} & d & K & L & \color{dred}{\,\,\,\,\,\,$b_t$\,\,\,\,\,} & d & K & L & \color{dred}{\,\,\,\,\,$b_t$\,\,\,\,} & d & K & L & \color{dred}{\,\,\,\,\,$b_t$\,\,\,\,}\\
\hline\hline
2 & 2 & 3 & \color{dred}{0.55} & 4 & 2 & 3 & \color{dred}{1.60} & 6 & 2 & 3 & \color{dred}{2.73} & 8 & 2 & 3 & \color{dred}{4.08} \\ \hline
2 & 2 & 4 & \color{dred}{0.53} & 4 & 2 & 4 & \color{dred}{1.48} & 6 & 2 & 4 & \color{dred}{2.57} & 8 & 2 & 4 & \color{dred}{3.57} \\ \hline
2 & 2 & 5 & \color{dred}{0.51} & 4 & 2 & 5 & \color{dred}{1.51} & 6 & 2 & 5 & \color{dred}{2.5} &  8 & 2 & 5 & \color{dred}{3.53} \\ \hline
2 & 2 & 6 & \color{dred}{\!\!0.499\!\!} & 4 & 2 & 6 & \!\!\!\color{dred}{1.4998}\!\!\! & 6 & 2 & 6 & \color{dred}{\!\!2.502\!\!} & 8 & 2 & 6 & \color{dred}{\!\!3.502\!\!} \\ \hline
2 & 3 & 2 & \color{dred}{0.86} & 4 & 3 & 2 & \color{dred}{2.26} & 6 & 3 & 2 & \color{dred}{3.96} & 8 & 3 & 2 & \color{dred}{5.54} \\ \hline
2 & 3 & 4 & \color{dred}{0.54} & 4 & 3 & 4 & \color{dred}{1.60} & 6 & 3 & 4 & \color{dred}{2.69} & 8 & 3 & 4 & \color{dred}{3.73} \\ \hline
2 & 3 & 5 & \color{dred}{\!\!0.51\!\!} & 4 & 3 & 5 & \!\!\color{dred}{1.53}\!\! & 6 & 3 & 5 & \color{dred}{\!\!2.566\!\!} & 8 & 3 & 5 & \color{dred}{\!\!3.593\!\!} \\ \hline
2 & 3 & 6 & \color{dred}{0.5} & 4 & 3 & 6 & \color{dred}{1.37} & 6 & 3 & 6 & \color{dred}{2.57} & 8 & 3 & 6 & \color{dred}{3.6} \\ \hline
2 & 4 & 2 & \color{dred}{0.7} & 4 & 4 & 2 & \color{dred}{2.4} & 6 & 4 & 2 & \color{dred}{4.9} & 8 & 4 & 2 & \color{dred}{6.3} \\ \hline
2 & 4 & 3 & \color{dred}{1.02} & 4 & 4 & 3 & \color{dred}{2.13} & 6 & 4 & 3 & \color{dred}{3.19} & 8 & 4 & 3 & \color{dred}{4.59} \\ \hline
2 & 4 & 5 & \color{dred}{0.51} & 4 & 4 & 5 & \color{dred}{1.60} & 6 & 4 & 5 & \color{dred}{2.63} & 8 & 4 & 5 & \color{dred}{3.71} \\ \hline
2 & 4 & 6 & \color{dred}{0.50} & 4 & 4 & 6 & \color{dred}{1.47} & 6 & 4 & 6 & \color{dred}{2.57} & 8 & 4 & 6 & \color{dred}{3.61} \\ \hline
2 & 5 & 2 & \color{dred}{0.78} & 4 & 5 & 2 & \color{dred}{2.48} & 6 & 5 & 2 & \color{dred}{4.4} & 8 & 5 & 2 & \color{dred}{7} \\ \hline
2 & 5 & 3 & \color{dred}{0.60} & 4 & 5 & 3 & \color{dred}{2.0} & 6 & 5 & 3 & \color{dred}{3.4} & 8 & 5 & 3 & \color{dred}{5.2} \\ \hline
2 & 5 & 4 & \color{dred}{0.85} & 4 & 5 & 4 & \color{dred}{2.04} & 6 & 5 & 4 & \color{dred}{3.26} & 8 & 5 & 4 & \color{dred}{4.15} \\ \hline
2 & 5 & 6 & \color{dred}{0.55} & 4 & 5 & 6 & \color{dred}{1.6} & 6 & 5 & 6 & \color{dred}{2.62} & 8 & 5 & 6 & \color{dred}{3.60} \\ \hline
2 & 6 & 2 & \color{dred}{\!\!0.862\!\!} & 4 & 6 & 2 & \color{dred}{\!\!2.583\!\!} & 6 & 6 & 2 & \color{dred}{4.4} & 8 & 6 & 2 & \color{dred}{8} \\ \hline
2 & 6 & 3 & \color{dred}{0.52} & 4 & 6 & 3 & \color{dred}{2.39} & 6 & 6 & 3 & \color{dred}{4.54} & 8 & 6 & 3 & \color{dred}{6} \\ \hline
2 & 6 & 4 & \color{dred}{0.63} & 4 & 6 & 4 & \color{dred}{1.79} & 6 & 6 & 4 & \color{dred}{3.04} & 8 & 6 & 4 & \color{dred}{4.35} \\ \hline
2 & 6 & 5 & \color{dred}{0.78} & 4 & 6 & 5 & \color{dred}{1.90} & 6 & 6 & 5 & \color{dred}{3.04} & 8 & 6 & 5 & \color{dred}{4.24} \\ \hline
\end{tabular}
\caption{Values of the numerical fits for $b_t(d,K,L)$ for $\Delta=3/2$ and $K\neq L$.}
\label{t.KneqL}
\end{table}
%


\section{Unequal blackening factors}\label{a.unequalblack}

As discussed in Section~\ref{ss.notime}, in the case $x\neq0$, one cannot relate Kasner exponents to the power of the first non-trivial singularity. One may attempt to circumvent the problem by assuming more general metrics. Consider black holes with unequal blackening factors
\begin{equation}\label{eq:GenMet}
      \dd s^2=-r^2f(r)\dd t^2+\frac{\dd r^2}{r^2 h(r)}+r^2\dd\vec{x}_{d-1}^{\,2},
\end{equation}
where now $f(r) = 1 - \sum_{j = 1}^{K} \mu_jr^{-2j}$ and $h(r)=1-\sum_{j=1}^{L}\nu_jr^{-2j}$ with $K\neq L$. 
In the near-singularity limit, the metric then takes on the Kasner form \eqref{eq:Kasner} with exponents
\begin{align}
    p_t = \frac{1-K}{L} \qq{and} p_i = \frac{1}{L}\,.
\end{align}
If one were to find the value of  $b_t(d,K,L)$ (or $b_x(d,K,L)$) in this setup, one could, in principle, disentangle the contributions from both $p_t$ and $p_i$.

To perform the holographic computations, we use
\begin{equation}\label{e.svist}
    f(r)=1-\frac{\mu}{r^{2K}}\qq{and} h(r)=1-\frac{\nu}{r^{2L}}.
\end{equation}
Setting $x=0$, the current sector can be written as
\begin{equation}
    G_T(\tau)=\tau^{-2\Delta}\sum_{n=0}^\infty\Lambda_n^{(L)}\tau^{2nN},
\end{equation}
where $N=\text{gcd}(K,L)$ and -- since the coefficient $b_t(d,K,L)$ seems to be independent of the concrete (non-zero) values of $\mu$ and $\nu$ -- we have set the parameters $\mu$ and $\nu$ to one. Using the generalised recurrence relation method one extracts $\Lambda^{(L)}_n$. Unlike in the $K=L$ case, here one in general finds cosine oscillations in the asymptotic form of the coefficients $\Lambda_n^{(L)}$:
\begin{equation}\label{e.KneqLasyfor}
    \Lambda^{(L)}_n\sim a^nn^{2\Delta-b_t(d,K,L)-1}\cos(cn+\phi),
\end{equation}
where the parameters $a$, $c$ and $\phi$ depend on $d$, $K$ and $L$. The cosine oscillations make it technically difficult to extract $b_t(d,K,L)$ -- and even more for the $b_x(d,K,L)$ for which the fits converge slower. In Table \ref{t.KneqL} we present the results of the fits for the coefficient $b_t(d,K,L)$.

With the current precision we were unable to determine the precise analytic form of $b_t(d,K,L)$ or $b_x(d,K,L)$. Nevertheless, we observe several important properties of $b_t(d,K,L)$:
First, whenever $K+L=8$, the cosine factor becomes constant or alternates as $(-1)^n$.  Second, whenever $K$ or $L$ equals 1, the coefficients $\Lambda_n^{(L)}$ do not admit the large-$n$ behaviour \eqref{e.KneqLasyfor}. Third, if a closed form formula for $b_t(d,K,L)$ exists, it has to satisfy $b_t(d,K,K)=\frac{d-1}{2}\frac{K}{K-\frac12}$.
We believe that with greater computational power, or more sophisticated fitting techniques, one should be able to obtain an explicit relation for $b_t(d,K,L)$ and $b_x(d,K,L)$.


\bibliographystyle{apsrev4-1}
\bibliography{draft}

\end{document}